\documentstyle[prl,aps,preprint]{revtex}
\begin{document}

\title{The incoherent part of the spin-wave polarization operator
 in the \mbox{\boldmath $t$-$J$\/} model}

\author {D. W. Murray and O. P. Sushkov$^{a}$}
\address { School of Physics, The University of New South Wales,\\
Sydney, 2052, Australia}

\maketitle

\begin{abstract}
A calculation of the spin-wave polarization operator is very important for
the analysis of the magnetic structure of high temperature superconductors.
We analyze the significance of the
incoherent part of the spin-wave polarization operator within the framework of
the $t$-$J$ model. This part is
calculated analytically for small doping with logarithmic accuracy.
We conclude that the incoherent part of the spin-wave polarization
operator is negligible in comparison with the coherent part.
\end{abstract}

\vspace{0.5cm}
\hspace{3.cm}keywords: antiferromagnetic order, spin fluctuations.
\vspace{1.cm}

\section{Introduction}
 It is  widely believed that the spin structure of the ground state
of high-temperature copper oxide superconductors is of the type of the Anderson
spin-liquid state\cite{And}. On the other hand it is well known that
at zero doping the materials under consideration are insulators
with long-range antiferromagnetic (AF) order. The instability of
long-range AF order under doping in the
$t$-$J$ model was probably first realized by Shraiman and Siggia\cite{Shr9}.
It was later demonstrated in numerous works (see e.g. refs.
\cite{Dom0,Sin0,Ede1,Iga2,Sus3}). The instability is due to the strong
interaction of spin-waves with mobile holes. The spin-wave polarization
operator is needed for the analysis of this
instability and especially for the derivation of the spin-liquid ground state.
The calculation of the so called
coherent part of the polarization operator is straightforward.
However there is also the incoherent part. The purpose of the present
work is to calculate this part and to provide a rigorous proof
of its smallness in comparison with the coherent part.
It means that in the analysis
of the magnetic properties one can neglect the incoherent
contribution to the spin-wave polarization operator. This
conclusion is in agreement with that of ref.\cite{Beck93}
based on the numerical computation.

\section{Hamiltonian}
The $t$-$J$ model is defined by the following Hamiltonian:
\begin{equation} \label{H}
  H = H_t + H_J
    = -t \sum_{<nm>\sigma} ( d_{n\sigma}^{\dag} d_{m\sigma} + \mbox{H.c.} )
     + J \sum_{<nm>}  {\bf S}_n \cdot {\bf S}_m,
\end{equation}
where $d_{n\sigma}^{\dag}$ is the creation operator of a hole with spin
 $\sigma$ ($\sigma= \uparrow, \downarrow$) at site $n$ on a
 two-dimensional square lattice.
The $d_{n\sigma}^{\dag}$ operators act in the Hilbert space
 with no double electron occupancy.
The spin operator is ${\bf S}_n = {1 \over 2} d_{n \alpha}^{\dag}
 ${\boldmath $\sigma$}$_{\alpha \beta} d_{n \beta}$.
$<nm>$ are pairs of nearest-neighbor sites on the lattice.
Below we set $J$ as well as the lattice spacing equal to unity.

At half-filling (one hole per site) the $t$-$J$ model is equivalent to
the Heisenberg AF model\cite{Hir5,Gro7} which has long-range
AF order in the ground state\cite{Oit1,Hus8}. Let us denote the
wave function of this ground state by $|0\rangle$.
This is the undoped system.
We consider the doped system based on the ground state
of the undoped system.
In spite of the destruction of the long-range AF order it is convenient to use
$|0\rangle$
and corresponding quasiparticle excitations as a basis set for
the doped system.
The effective Hamiltonian for the $t$-$J$ model was derived in terms of
these quasiparticles in the papers\cite{Suhf,Cher3,Kuch3}:
\begin{equation}
\label{Heff}
H_{eff}=\sum_{{\bf k}\sigma}\epsilon_{\bf k}h_{{\bf k}\sigma}^{\dag}
h_{{\bf k}\sigma}+
\sum_{\bf q}\omega_{\bf q}(\alpha_{\bf q}^{\dag}\alpha_{\bf q}
+\beta_{\bf q}^{\dag}\beta_{\bf q}) +H_{h,sw} + H_{hh}.
\end{equation}
It is expressed in terms of normal spin-waves on an AF
background $\alpha_{{\bf q}}, \beta_{{\bf q}}$
(see e.g. review\cite{Manousakis}) and
composite hole operators $h_{{\bf k}\sigma}$ ($\sigma = \pm 1/2$).
The summations over ${\bf k}$ and ${\bf q}$ are restricted inside the
Brillouin zone of one sublattice where
$\gamma_{\bf q}= {1\over 2} (\cos q_x+\cos q_y)\ge 0$.
The spin-wave dispersion relation is
\begin{equation}
\label{swdisp}
\omega_{\bf q}=2\sqrt{1-\gamma_{\bf q}^2}\ , \ \ \ \ \
\omega_{\bf q} \approx \sqrt{2}|{\bf q}| \ \mbox{for} \ q \ll 1.
\end{equation}

The properties of single holes are well established (for a review see
ref.\cite{Dag4}). The wave function of a single hole can be represented as
 $\psi_{{\bf k}\sigma} = h_{{\bf k}\sigma}^{\dag} |0\rangle$.
At large $t$ the composite hole operator $h_{{\bf k}\sigma}^{\dag}$ has
a complex structure. For example at $t/J=3$ the weight of the bare hole in
$\psi_{{\bf k} \sigma}$ is about 25\%, the weight of configurations of the
type ``bare hole + 1 magnon'' is $\sim$50\% and for configurations of the
type ``bare hole + 2 or more magnons'' it is $\sim$25\%. The dressed hole is a
normal
fermion. The hole energy $\epsilon_{\bf k}$
has minima at ${\bf k}={\bf k}_0$, where
${\bf k}_0=(\pm \pi/2, \pm\pi/2)$. For $t \le 5$ the dispersion can be
well approximated by the expression\cite{Sus2}
\begin{equation}
\label{hdisp}
\epsilon_{\bf k} \approx E_0+2-\sqrt{0.66^2+4.56t^2-2.8t^2\gamma_{\bf k}^2}
   + {1\over4} \beta_2 (\cos k_x -\cos k_y)^2.
\end{equation}
The numerical values in this formula are some combinations of the
Heisenberg model spin correlators. The constant $E_0$ defines a reference
level for the energy. To find $\epsilon_{\bf k}$ with respect to the
undoped system one has to set $E_0=0$. However for the present work it
is convenient to set $\epsilon_{\bf k_0}=0$ and therefore
$E_0=\sqrt{0.66^2+4.56t^2}-2$.
The coefficient $\beta_2$ is small and therefore the dispersion is almost
degenerate along the face of the magnetic Brillouin zone $\gamma_{\bf k}=0$.
According to refs. \cite{Mart1,Giam3} $\beta_2 \approx 0.1\,t$
for $t \ge 0.33$.
To avoid misunderstanding we note that formula (\ref{hdisp})
is not valid for very large $t$ where the hole band width is saturated
at the value of the order of unity and does not increase with $t$.
However, physically we are interested in $t \approx 3$ (see e.g.
refs.\cite{Esk0,Fla1,BCh3}) where (\ref{hdisp}) works well.
Near the band minima, ${\bf k}_0$ the dispersion (\ref{hdisp}) can be
presented in the usual quadratic form:
\begin{equation}
\label{hdisp1}
\epsilon_{\bf p} \approx {1\over2} \beta_1 p_1^2 + {1\over2} \beta_2 p_2^2,
 \hspace{1.0cm} \beta_2 \ll \beta_1,
\end{equation}
where $p_1$ ($p_2$) is the projection of ${\bf k}-{\bf k}_0$
 on the direction orthogonal (parallel) to the face of the
 magnetic Brillouin zone (fig. 1). From eq. (\ref{hdisp}) for $t \gg 0.33$
we find $\beta_1 \approx 0.65 t$, hence the mass anisotropy is
$\beta_1/\beta_2 \approx 7$.
For small hole concentrations, i.e. $\delta \ll 1$, the holes are localized in
momentum space in the vicinity of the minima of the band
${\bf k}_0 = (\pm \pi/2, \pm \pi/2)$ and the Fermi surface
consists of ellipses.
(We recall that $\delta=0$ corresponds to an insulator with long-range
AF order.) The Fermi energy and Fermi momentum of non-interacting holes are
\begin{equation} \label{ep}
 \epsilon_F \approx \frac{1}{2} \pi (\beta_1 \beta_2)^{1/2} \delta \qquad
\mbox{and} \qquad p_F \approx \sqrt{p_{1F}p_{2F}} \approx (\pi \delta)^{1/2}.
\end{equation}
The Fermi momentum $p_F$ is measured from the center of
the corresponding ellipse.

The interaction of a composite hole with  spin-waves
is of the form (see e.g. refs.\cite{Mart1,Liu2,Suhf})
\begin{eqnarray}
\label{hsw}
&& H_{h,sw} = \sum_{\bf k,q} g_{\bf k,q}
    \biggl(
  h_{{\bf k}+{\bf q}\downarrow}^{\dag} h_{{\bf k}\uparrow} \alpha_{\bf q}
  + h_{{\bf k}+{\bf q}\uparrow}^{\dag} h_{{\bf k}\downarrow} \beta_{\bf q}
  + \mbox{H.c.}   \biggr),\\
&& g_{\bf k,q} = 2\sqrt{2}f
 (\gamma_{\bf k} U_{\bf q} + \gamma_{{\bf k}+{\bf q}} V_{\bf q})
\to 2^{1/4}f\sqrt{{1\over{q}}}
(q_x \sin k_x + q_y \sin k_y) \ \mbox{for} \ q \ll 1.
\nonumber
\end{eqnarray}
Here $U_{\bf q}=\sqrt{{1\over{\omega_{\bf q}}}+{1\over 2}}$ and
$V_{\bf q}=-\mbox{sign}(\gamma_{\bf q})\sqrt{{1\over{\omega_{\bf q}}}-{1\over
2}}$
are the parameters of the Bogoliubov transformation diagonalizing the spin-wave
Hamiltonian. The hole spin-wave coupling constant $f$ is a function of $t$
evaluated in the work\cite{Suhf}.
For large $t$ the coupling constant is $t$-independent, $f \approx 2$.
Let us stress that even for $t > J$ the quasihole-spin-wave
interaction (\ref{hsw}) has the same form as
for $t \ll J$ (i.e. as for bare hole  operators)
with an added renormalization factor (of the order $J/t$ for $t \gg J$).
This remarkable property of the $t$-$J$ model is due to the absence
 of a single loop correction to the hole-spin-wave vertex.
It was first stated implicitly by Kane, Lee and Read\cite{Kan9}.
In refs.\cite{Mart1,Liu2,Suhf} it was
explicitly demonstrated that the vertex corrections with different kinematic
structure are of the order of a few percent at $t/J \approx 3$.
There is also some $q$-dependence of the coupling constant $f$.
For example $f(q=\pi)\approx 1.15f(q=0)$ at $t/J = 3$ (see refs.
\cite{Kuch3,Bel95}). However this dependence is weak and is
beyond the accuracy of the calculation of the renormalized value of $f$.
Therefore we neglect this dependence.

Finally there is the contact hole-hole interaction $H_{hh}$ in the
effective Hamiltonian (\ref{Heff}). $H_{hh}$ is discussed in detail in
refs.\cite{Cher3,Kuch3,Bel95}. It is proportional to some function
$A(t)$. For small $t$ this function approaches the value $-0.25$, which
gives the well known hole-hole attraction induced by the reduction of the
number of missing antiferromagnetic links. However for
realistic superconductors $t\approx 3$ (see e.g. refs.\cite{Esk0,Fla1,BCh3}).
Surprisingly the function $A(t)$ vanishes exactly at $t\approx 3$ and this
means that the mechanism of contact hole-hole attraction is switched off.
In contrast the spin-wave exchange mechanism $H_{h,sw}$ is negligible
for small $t$ for which $f \sim t$ and it is most important for
large $t$ for which $f$ approaches 2. We are interested
in ``physical'' values of $t$: $t \approx 3$. Therefore in the present
work we neglect the contact interaction ($H_{hh}=0$) and consider only
the hole-spin-wave interaction at $t=3$. The corresponding value of $f$
according to\cite{Suhf} is $f=1.8$. Thus all numerical values in
the present work are calculated at $t=3$, $\beta_1/\beta_2=7$ and
$f=1.8$.

\section{Polarization Operator}
There are two types of spin-waves in the $t$-$J$ model, $\alpha$
and $\beta$. Therefore, one generally has to introduce a set of
spin-wave Green's functions\cite{Iga22,Khal3}: $D_{\alpha \alpha}$,
$D_{\alpha \beta}$, $D_{\beta \alpha}$, and $D_{\beta \beta}$.
However if we restrict ourselves to the long wavelength limit,
the consideration can be simplified.
It is well known that in the long wavelength limit the Heisenberg model is
equivalent to the nonlinear $\sigma$-model (see e.g. review
paper\cite{Manousakis}). Therefore the usual field theory crossing symmetry
is valid. The same is valid for the $t$-$J$ model. This is evident
from eq. (\ref{hsw}): at $q \ll 1$ the
vertex $g_{\bf k,q} \approx g_{\bf k-q,q}$.
This means that instead of considering a set of Green's functions
one can introduce a single combined Green's function of vector excitation:
\begin{equation}
\label{GF}
D(\omega,{\bf q})={{2\omega_{\bf q}}\over{\omega^2-\omega_{\bf q}^2-
2\omega_{\bf q}\Pi(\omega,{\bf q})}},
\end{equation}
where $\Pi(\omega,{\bf q})$ (or $P(\omega,{\bf q})$) is the mobile holes
polarization operator.
For the system to be stable the condition
\begin{equation}
\label{stab}
\omega_{\bf q} + 2 \Pi(0,{\bf q}) > 0
\end{equation}
should be fulfilled. Otherwise the Green's function (\ref{GF}) would
possess poles with imaginary $\omega$.

In the single loop approximation
for dressed quasiholes the polarization operator is given by the
diagram in fig. 2. The quasihole Green's function in this diagram is determined
by the effective Hamiltonian (\ref{Heff}). Let us call this
contribution to the polarization operator the ``coherent part''.
The meaning of this term is elucidated below.
In this paper we neglect the
interaction between quasiholes and consider them as a normal
Fermi-liquid. In this approximation the calculation of a single loop
polarization operator is very simple. In the static long wavelength
limit the result is\cite{Sus3}
\begin{equation}
\label{P0q}
\Pi_{coh}(0,{\bf q})\approx -{{\sqrt{2}f^2}\over{\pi\sqrt{\beta_1 \beta_2}}}q
\ \ \ \mbox{for} \ \ \ q \ll p_F.
\end{equation}
This is independent of the hole concentration! Using this result one can
easily prove that the stability condition (\ref{stab}) is violated because
$-2\Pi(0,{\bf q})/\omega_{\bf q}=
2f^2/(\pi \sqrt{\beta_1 \beta_2}) \approx 2.8$.
This is a well known fact concerning the instability of long-range AF order
under doping (see e.g. refs.\cite{Shr9,Dom0,Sin0,Ede1,Iga2,Sus3}).

Let us now explain the obtained result in terms of bare $d_{n\sigma}$
which enter into the bare Hamiltonian (\ref{H}).
The Green's function of a bare hole is of the form
\begin{equation}
\label{hGF}
G(\epsilon,{\bf k})={{Z}\over{\epsilon - \epsilon_{\bf k}+i0}} + G_{incoh},
\end{equation}
where $Z$ is the quasiparticle residue. If we use the bare hole Green's
function we also have to use the bare hole-spin-wave coupling constant
$f_{bare}=2t$ in the vertex $g_{\bf k,q}$ (\ref{hsw}). Substitution
of the pole (coherent) part of (\ref{hGF}) into the single loop polarization
operator gives the combination $f_{bare}Z$, but this is exactly the effective
coupling constant $f$\cite{Suhf}.
Thus in the single loop approximation the effective theory with
Hamiltonian (\ref{Heff}) only takes into account the coherent (pole)
part of the bare hole Green's function (\ref{hGF}). Because of this
the contribution (\ref{P0q}) is called the coherent part.

The incoherent part of the polarization operator which comes from $G_{incoh}$
in (\ref{hGF}) is proportional to the hole concentration $\delta$.
Therefore it is
negligible for $\delta \ll 1$. This conclusion agrees with that of
Becker and Muschelknautz\cite{Beck93}.

Now we will calculate $P_{incoh}$ explicitly.
We give two different calculations of $P_{incoh}$. The first calculation
is based on the self-energy approach. It is well known that the
self-energy as well as the polarization operator is in essence an energy
shift of the system. Let us consider the diagram in fig. 3a which represents
the energy shift of a hole (solid line) in an external spin-wave
(wavy line). If we join up the wavy lines and integrate over the momenta of the
spin-wave we get the usual hole self energy (fig. 3b) which is already
taken into account in the correct (renormalized) hole dispersion
(\ref{hdisp}) and the renormalized value of the
spin-wave coupling constant $f$.
If we join up the solid lines and integrate over the momenta of the hole
we get the coherent polarization operator (fig. 3c) which has also
already been taken into account. Let us consider now the first correction to
the diagram fig. 3a. It is given by the diagrams in fig. 4. (We remind the
reader that there is no one loop correction to the spin-wave vertex.)
This is the correction to the hole energy in an external spin-wave
or the correction to the energy of the spin-wave in the presence of a
single hole. The calculation of this correction is very simple if one
uses Schr\"{o}dinger perturbation theory. Consider for example the case
when the external spin-wave is of the $\beta$-type, i.e. for this wave
the z-projection of spin is $S_z=+1$. The arrows in fig. 4 indicate
the projections of hole spin (note that there is a spin flip at
each vertex). It is evident that for spin up
diagram 4b vanishes and similarly for spin down diagram 4a vanishes. We will
see that the integral over $Q$ is logarithmically divergent. Therefore
we set $q \ll Q \ll 1$. Near the band minimum the hole energy is quadratic
in momenta (\ref{hdisp1}). Therefore it can be neglected in comparison
with the energy of the intermediate spin-wave $\omega_{\bf Q}$ which
is linear in momentum. This means that the energies of all three
intermediate states in fig. 4a as well as in fig. 4b are approximately
equal to $\omega_{\bf Q}$. Therefore the energy correction corresponding
to fig. 4 is independent of the hole spin projection and equals
\begin{equation}
\label{deps}
\delta \epsilon = -g^2_{{\bf k},{\bf q}}\sum_{\bf Q}
{{g^2_{{\bf k},{\bf Q}}}\over{\omega_{\bf Q}^3}}
=-{{f^4}\over{\sqrt{2}q}}(q_x \sin k_x +q_y \sin k_y)^2
\int {{(Q_x \sin k_x +Q_y \sin k_y)^2}\over{Q^4}} {{d^2 Q}\over{(2\pi)^2}}.
\end{equation}
The integration over $Q$ is restricted by the limits $Q_{min} \sim p_F
=\sqrt{\pi \delta}$ and $Q_{max} \sim \pi$.
The integral over $Q$ is logarithmically divergent and therefore it
gives a big logarithm: $\ln (Q_{max}/Q_{min}) \approx 0.5\ln (1/\delta)$.
When the momentum of the hole is close to the band minimum,
${\bf k} \approx {\bf k}_0=(\pi/2, \pm \pi/2)$, one gets
\begin{equation}
\label{deps1}
\delta \epsilon = {{f^4}\over{4\sqrt{2}\pi}}
{{(q_x \pm q_y)^2}\over{q}} \ln{\delta}.
\end{equation}
To find the incoherent spin-wave polarization operator we now need
to sum (\ref{deps1}) over all occupied hole states (i.e. sum
over ${\bf k}$). Thus for $\omega \ll \epsilon_F$ and $q \ll p_F$ the result is
\begin{equation}
\label{Pincoh}
P_{incoh}(\omega,{\bf q}) \approx -{{f^4}\over{4\sqrt{2}\pi}} \cdot \delta
\cdot \ln{{1}\over{\delta}} \cdot q.
\end{equation}
There are also higher order diagrams which contribute to $P_{incoh}$.
An example is presented in fig. 5. However all of these diagrams do
not contain the big logarithm ($\ln(1/\delta)$) and therefore can
be neglected according to the accepted accuracy.

To elucidate the meaning of $P_{incoh}$ we will now give a different
derivation of (\ref{Pincoh}). The first correction to the single loop
polarization operator arises from the diagram presented in fig. 6.
After integration over the energies in the closed loops one
gets the following expression for the contribution of this diagram:
\begin{eqnarray}
\label{P6}
&&P_6(\omega,{\bf q})=\int {{d^2 k}\over{(2\pi)^2}}{{d^2 Q}\over{(2\pi)^2}}
g^2_{{\bf k},{\bf Q}}g^2_{{\bf k},{\bf q}}\times\\
&&\biggl({{n_{\bf k-q}(1-n_{\bf k-Q})-n_{\bf k}(1-n_{\bf k-Q})}\over
{(\epsilon_{\bf k-q}-\epsilon_{\bf k}+\omega)^2
(\epsilon_{\bf k-q}-\epsilon_{\bf k-Q}-\omega_{\bf Q}+\omega)}}+
{{n_{\bf k-q}n_{\bf k-Q}-n_{\bf k}n_{\bf k-Q}}\over
{(\epsilon_{\bf k-q}-\epsilon_{\bf k}+\omega)^2
(\epsilon_{\bf k-q}-\epsilon_{\bf k-Q}+\omega_{\bf Q}+\omega)}}+\nonumber \\
&&{{-n_{\bf k}(1-n_{\bf k-Q})}\over
{(\epsilon_{\bf k-Q}-\epsilon_{\bf k}+\omega_{\bf Q})^2
(\epsilon_{\bf k-Q}-\epsilon_{\bf k-q}+\omega_{\bf Q}-\omega)}}+
{{(1-n_{\bf k})n_{\bf k-Q}}\over
{(\epsilon_{\bf k-Q}-\epsilon_{\bf k}-\omega_{\bf Q})^2
(\epsilon_{\bf k-Q}-\epsilon_{\bf k-q}-\omega_{\bf Q}-\omega)}}
\biggr),\nonumber
\end{eqnarray}
where $n_{\bf k}$ is the ground state occupation number. The first two terms
in the expression (\ref{P6}) have second powers of
$(\epsilon_{\bf k-q}-\epsilon_{\bf k}+\omega)$ in the denominator.
This factor is exactly the denominator of the one-loop
polarization operator fig. 2. It means that the first two terms correspond
to the renormalization of the pole (coherent) part of the hole Green's
function (\ref{hGF}). They are actually already taken into account in
$P_{coh}$ because for the calculation of $P_{coh}$ we have used the
dressed hole Green's function which absorbs all of the corrections to the
pole part. Thus the new contribution, corresponding to $P_{incoh}$,
arises from the last two terms
in (\ref{P6}). After carrying out some algebraic manipulation and
using the fact that $q$ is small (and so, for example
$\epsilon_{\bf k-q} \approx \epsilon_{\bf k}$) $P_{incoh}$ can be
written as
\begin{equation}
\label{Pincoh1}
P_{incoh}(0,{\bf q})=-2\int {{d^2 k}\over{(2\pi)^2}}{{d^2 Q}\over{(2\pi)^2}}
g^2_{{\bf k},{\bf Q}}g^2_{{\bf k},{\bf q}}
{{n_{\bf k}(1-n_{\bf k-Q})}\over
{(\omega_{\bf Q}+\epsilon_{\bf k-Q}-\epsilon_{\bf k})^3}}.
\end{equation}
It is evident that this expression is equivalent to
(\ref{deps}). Neglecting $\epsilon_{\bf k-Q}-\epsilon_{\bf k}$ in
comparison with $\omega_{\bf Q}$ and carrying out the integration
over ${\bf Q}$ with logarithmic accuracy we once more come to the
result (\ref{Pincoh}) for $P_{incoh}$.

\section{Conclusion}
In the present work we have analytically calculated the incoherent
contribution to the spin-wave polarization operator. This analytical
calculation is possible because of the logarithmic enhancement
of the corresponding diagrams. The ratio of the incoherent contribution
(\ref{Pincoh}) to the coherent one (\ref{P0q}) is
\begin{equation}
\label{ratio}
{{P_{incoh}(0,{\bf q})}\over{P_{coh}(0,{\bf q})}}=
{1\over{8}}f^2 \sqrt{\beta_1 \beta_2} \cdot \delta \cdot
\ln{{1}\over{\delta}} \ .
\end{equation}
For the parameters of the $t$-$J$ model corresponding to realistic
high-T$_c$ superconductors and for hole concentration $\delta \le 0.2$
this ratio does not exceed 10\%. It means that in the analysis
of the magnetic properties one can neglect the incoherent
contribution to the spin-wave polarization operator.

{\bf FIGURE CAPTIONS}
\\
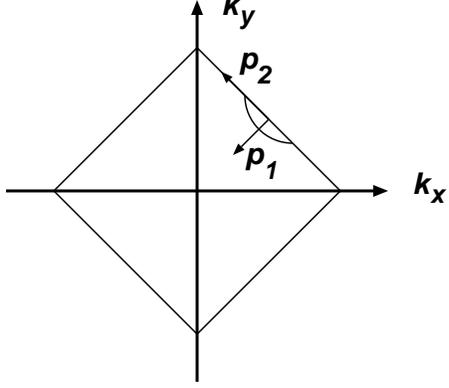
FIG. 1. The magnetic Brillouin zone.\\
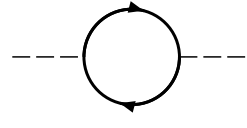
FIG. 2. The coherent spin-wave polarization operator.\\
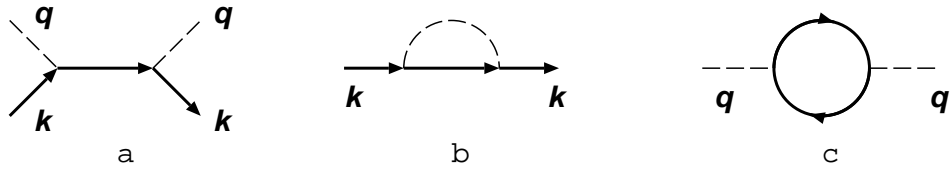
Fig. 3. a) The leading order contribution to the energy shift of a hole (solid
line) in
an external spin-wave (dashed line). b) The hole self-energy.
c)The coherent spin-wave polarization operator.\\
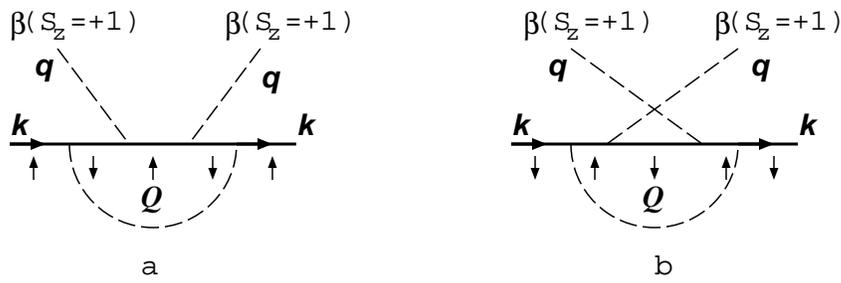
FIG. 4. The fourth order correction to the energy of a hole in an
external spin-wave.\\
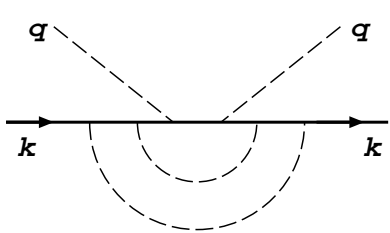
FIG. 5. A typical fifth order diagram for the energy of a hole in an
external spin-wave.\\
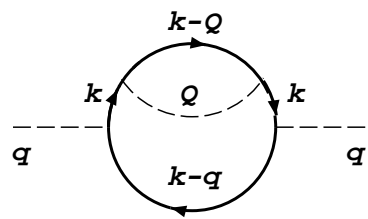
FIG. 6. The two loop correction to the spin-wave polarization operator.

\end{document}